\documentclass{acmart}

\AtBeginDocument{%
  \providecommand\BibTeX{{%
    \normalfont B\kern-0.5em{\scshape i\kern-0.25em b}\kern-0.8em\TeX}}}

\setcopyright{acmlicensed}
\copyrightyear{2018}
\acmYear{2018}
\acmDOI{XXXXXXX.XXXXXXX}

\acmConference[UMAP '24]{User Modelling and User-Adapted Interaction}{July, 1-4, 2024}{Cagliari, Italy}
\acmISBN{978-1-4503-XXXX-X/18/06}

\usepackage{researchpack}
\usepackage[capitalize,nameinlink]{cleveref}
\usepackage{soul}

\begin{document}

\title[Preference Elicitation in Interactive and User-centered Algorithmic Recourse: an Initial Exploration]{Exploiting Preference Elicitation in Interactive and User-centered Algorithmic Recourse: An Initial Exploration}

\author{Seyedehdelaram Esfahani}
\author{Giovanni De Toni} 
\author{Bruno Lepri}
\author{Andrea Passerini}
\author{Katya Tentori} 
\author{Massimo Zancanaro}

\renewcommand{\shortauthors}{Esfahani et al.}

\begin{abstract}

Algorithmic Recourse aims to provide actionable explanations, or recourse plans, to overturn potentially unfavourable decisions taken by automated machine learning models. In this paper, we propose an interaction paradigm based on a \textit{guided} interaction pattern aimed at both eliciting the users' preferences and heading them toward effective recourse interventions. In a fictional task of money lending, we compare this approach with an \textit{exploratory} interaction pattern based on a combination of alternative plans and the possibility of freely changing the configurations by the users themselves. Our results suggest that users may recognize that the \textit{guided} interaction paradigm improves efficiency. However, they also feel less freedom to experiment with "what-if" scenarios. Nevertheless, the time spent on the purely \textit{exploratory} interface tends to be perceived as a lack of efficiency, which reduces attractiveness, perspicuity, and dependability. Conversely, for the \textit{guided} interface, more time on the interface seems to increase its attractiveness, perspicuity, and dependability while not impacting the perceived efficiency. That might suggest that this type of interfaces should combine these two approaches by trying to support \textit{exploratory} behavior while gently pushing toward a \textit{guided} effective solution.
\end{abstract}



\maketitle

\section{Introduction}

Alice is applying for a loan at a respected bank in her hometown that uses an automated machine learning system to decide whether to accept or reject loan applicants. Unfortunately, Alice gets denied the loan and the bank cannot provide her with any explanation about their decision, since their system is a "black-box": it performs well, but it is too complicated for humans to understand.
Alice's story is fictional, but there are already examples of machine learning systems applied to decision-making tasks with a high impact on human lives, such as credit scoring \cite{bhatore2020machine} and recidivism prediction \cite{wang2023pursuit}. 

Indeed, more than just an explanation about the reasons for the rejection of her request, Alice would benefit from practical suggestions on how to reverse the algorithm decision. \textit{Algorithmic recourse} (AR) \cite{wachter_counterfactual_2017, karimi_survey_2021} is a particular case of explanations based on counterfactual analysis \cite{guidotti_counterfactual_2022} that can provide a sequence of actions capable of overturning an unfavorable machine-driven decision \cite{ustun_actionable_2019}. For example, a recourse suggestion would provide Alice with the actions she needs to take to obtain a loan (e.g., "increase your monthly salary by 100\$").

In this respect, algorithmic recourse can better respond to the need for transparency and fairness required for AI-based systems by several national authorities in a context where AI-based tools can significantly impact people's lives
\cite{wachter_counterfactual_2017}. This aspect is gaining increasing importance as one effective way to implement the rights sought by the European GDPR \cite{voigt2017}.

Although there is a growing body of research on this topic \cite{karimi_survey_2021, verma2020counterfactual}, the assessment of the state-of-the-art recourse algorithms is usually based on quantitative benchmarking with metrics such as validity, minimality or similarity
\cite{bodria_benchmarking_2023}. While helpful in appreciating the potential of algorithmic recourse strategies, these approaches fall short of assessing the effectiveness of recourse strategies for real users.

While an algorithm must be efficient and effective, it is also of paramount importance that the interaction paradigm sustains and fosters empowerment by providing a creative space of opportunities to explore while limiting confusion and false expectations.
In this direction, the recent work by Wang and colleagues \cite{wang_gam_2023} proposes an algorithm based on integer linear programming to generate counterfactual explanations, together with a graphical user interface called GAM Coach, to enable end users to develop recourse plans iteratively. The authors refer to their approach as an \textit{exploratory interface} by citing the seminal work by Shneiderman on Human-centered AI Systems \cite{shneiderman_bridging_2020, shneiderman2022human} since it allows users to freely manipulate the values of the features involved in the decision process.

An evaluation study of GAM Coach provided helpful insights on designing user interfaces to support users in leveraging these algorithms effectively. Nevertheless, the authors noted how the exploration might become complicated and require several interactions to learn how it works. As a design lesson, they suggest that “developers can use the log data [...] to train a new model to predict users’ preference configurations” \cite{wang_gam_2023}. Another interesting result of that study is that, although participants appreciate being able to control recourse generation with what-if questions, “interactivity and transparency could occasionally confuse users with counterintuitive recourse plans”. \cite{wang_gam_2023}.

Indeed, a less explored but critical desideratum is the degree of \textit{personalization} of the recourse plans to match the users' needs. Tailoring recourse to users requires an \textit{interactive} strategy to include humans in the generation process since some users' preferences cannot be inferred by their features alone \cite{detoni2024personalized}.

Our work investigates a different interaction paradigm with respect to the \textit{purely exploratory} approach of GAM Coach, while still aimed at implementing the \textit{proactive principle} proposed for Human-centered AI Systems: "to probe the algorithm boundaries with different inputs" \cite{shneiderman_bridging_2020}. Specifically, we use an algorithmic recourse method enriched with preference elicitation capabilities which realizes the \textit{personalization} of a recourse plan. We propose an approach based on a \textit{guided-interaction} pattern aimed at directly eliciting the user’s preferences needed by the algorithm to head the user toward a satisfying recourse intervention quickly. 

We believe that limiting the exploratory opportunities within an optimized \textit{step-by-step} process may help relieve the user of the possible overwhelming amount of choices in a typical recourse negotiation while still preserving the gist of the prospective strategy by providing users with "a better understanding of each step in the process so that they can prevent mistakes" \cite{shneiderman2022human}. In this regard, our primary research question concerns the user experience of a novel interaction strategy that we refer to as \textit{guided interaction}. This strategy maintains an exploratory interaction style while establishing boundaries via an optimized preference elicitation process.

\begin{quote}    
\textit{RQ:} To what extent do users recognize and appreciate a \textit{guided-interaction} strategy  as measured by the pragmatic and hedonic dimensions of their overall experience?
\end{quote}

To provide a first answer to our research question, we present the design of a graphical user interface based on the \textit{guided-interaction} pattern as the foreground of a novel preference elicitation algorithm for recourse. Then, we compare it with a pilot study in a typical scenario for recourse problems (a fictional task of money lending) to a \textit{purely exploratory} interface inspired by the GAM Coach one \cite{wang_gam_2023}. 
 
\section{Related Work}

Algorithmic Recourse is concerned with providing actionable suggestions, or \textit{interventions}, that show how to overturn an unfavorable prediction of a machine learning model via counterfactual scenarios \cite{wachter_counterfactual_2017, byrne2019counterfactuals}.
Formally, let a user sampled from a distribution $P(X)$ be $\vx = \{x_1, \ldots, x_d\} \in \calX$, where $x_i$ is a single feature (\eg age, education level, etc.). Each user $x$ is assigned a binary outcome $y \in \{0,1\}$ following $P(Y \mid X)$.
Given a dataset $\{(x, y)_i\}_{i=1}^N$ sampled from $P(X,Y)$, we can learn a classifier $h: \calX \rightarrow \{0,1\}$ which predicts the class $y$ given a user $x$.
Following our previous example, $y=1$ could mean "$x$ will repay the loan".
We also define the (possibly unknown) user preferences as a vector $\vw \in \calW$ sampled from a distribution $P(W)$.
An intervention is a sequence of actions $I = \{ a^{(1)}, \ldots, a^{(N)}\}$ where each action $a^{(i)} \in \calA$ suggests a modification to a single \textit{actionable} feature \eg "reduce your monthly spending by 100\$". 
Given the classifier $h: \calX \rightarrow \{0, 1\}$, in algorithmic recourse we aim at finding the best sequence of actions $I$ optimizing the following objective:
\begin{equation}
    I^* = \argmin_I \; c(I, \vx, \vw) \quad \mathrm{s.t.} \quad h(I(\vx)) \neq h(\vx)
    \label{eqn:algorithmic-recourse-objective}
\end{equation}
where $c: \calI \times \calX \times \calW \rightarrow \bbR^+$ is a cost function. In practice, the idea is to find the \textit{cheapest} intervention following the user preferences that, if applied to the initial profile, will induce a different outcome by the classifier.
Optimizing \cref{eqn:algorithmic-recourse-objective} means solving both a \textit{NP-hard} combinatorial problem together with estimating the user preferences.
There is already a plethora of methods for optimizing \cref{eqn:algorithmic-recourse-objective} \cite{karimi_survey_2021, verma2020counterfactual}, which however discard user preferences or make some unrealistic assumptions.
Many algorithmic recourse approaches assume preferences are fully specified beforehand \cite{wachter_counterfactual_2017, ramakrishnan2020synthesizing, karimi2020algorithmic, detoni2023synthesizing} or they assume users can define them via numerical constraints \cite{poyiadzi2020face} or by quantifying the cost of each action \cite{mahajan2019preserving}. However, it is well known that it is hard for users to specify preferences quantitatively \cite{Keeney1993}. 

For what concerns user interaction and assessment of user experience, recent works have explored ways to revive users' agency when building counterfactual examples to understand machine learning models, such as ViCE \cite{gomez2020vice} and DECE \cite{cheng2021dece}. However, to the best of our knowledge, there are no other proposals, beyond GAM Coach \cite{wang_gam_2023}, considering both algorithmic aspects and users' interaction aspects.

\section{Structuring Interaction with Recourse based on Preference Elicitation}

On the side of the algorithmic recourse, our approach is inspired by the algorithm proposed by De Toni and colleagues \cite{detoni2024personalized} which integrates recourse with Preference Elicitation \cite{boutilier2002pomdp} to interactively estimate user preferences via pairwise comparison queries.
While a detailed description and a formal evaluation of the algorithm are beyond the scope of this short paper, we present the gist of the algorithm to facilitate the understanding of the proposed \textit{guided-interaction} paradigm.

Our approach is based on the idea of eliciting user preferences via an interaction protocol that asks the user to choose the preferred intervention from a small set $k$ of alternatives. The options are selected and proposed to the user at each algorithm interaction. We use a greedy approach to select the alternatives maximizing the information gain \cite{viappiani2010optimal}. The user choice is then used to update a distribution over cost weights P(W), progressively converging towards the true user preferences $\vw^*$ used by \cref{eqn:algorithmic-recourse-objective}. 
Lastly, this preference elicitation strategy is integrated into a reinforcement learning agent coupled with a discrete search procedure (Monte Carlo Tree Search \cite{coulom2006efficient}) to efficiently discover recourse solutions.
This approach realizes an instance of the coactive learning strategy as proposed by \cite{ShivaswamyJ15}.

\begin{figure}[t]
  \centering
  \includegraphics[width=0.45\textwidth]{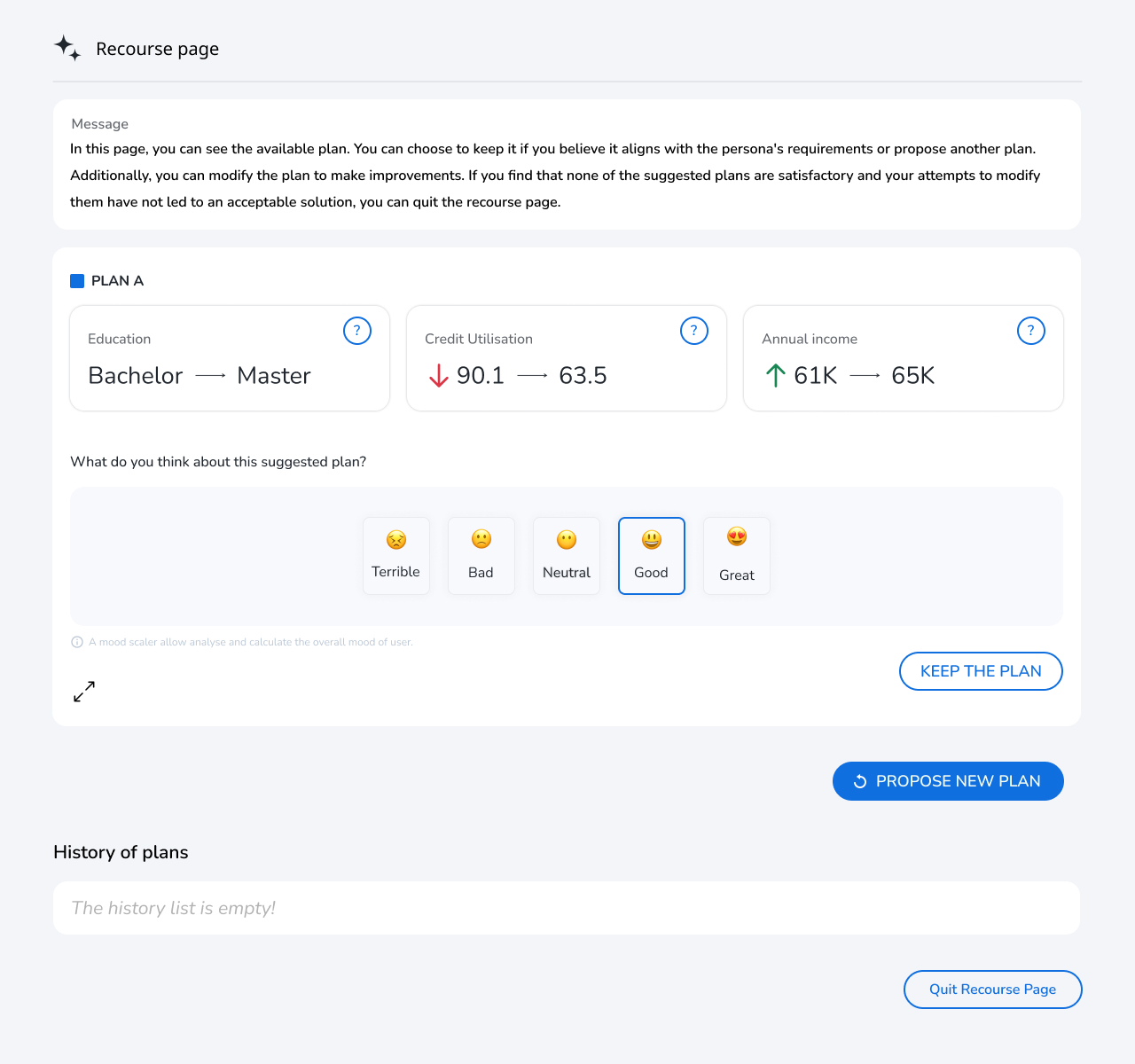}
  \includegraphics[width=0.45\textwidth]{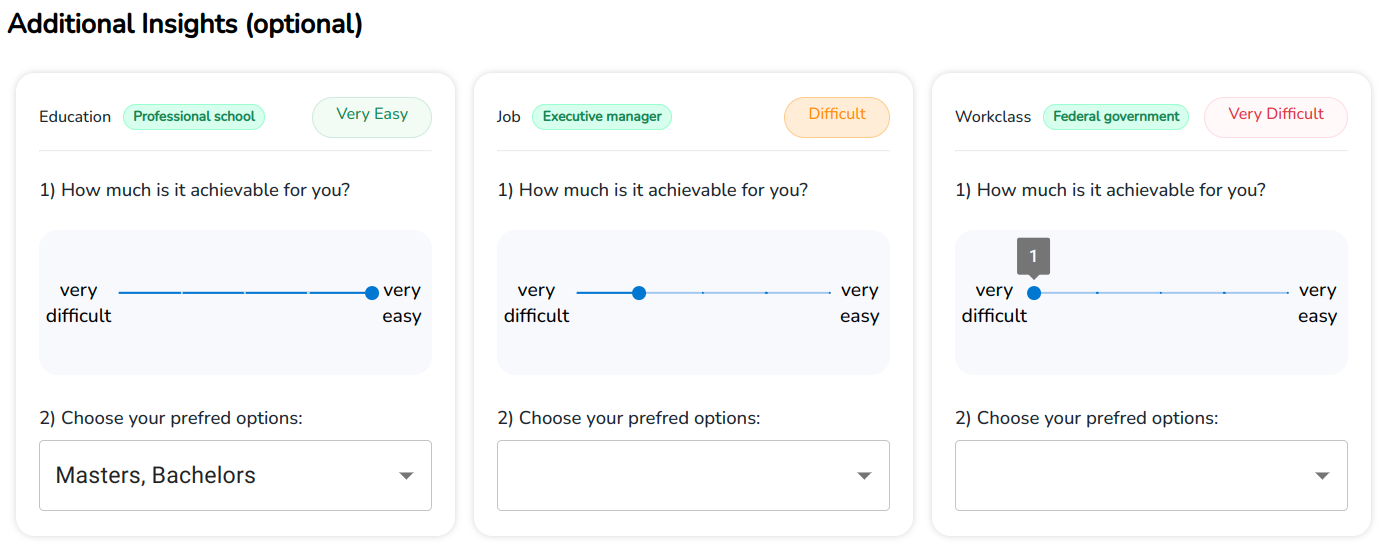}
  \caption{The \textit{guided} interface that uses the algorithmic recourse method enriched with preference elicitation: the user can only express preferences on the proposed features since they have been selected by the algorithm as those that optimize the learning of a personalized cost function.}
  \label{fig:PrefElic_figure}
  \label{fig:guided-interface-control-tab}
  \Description{On the left, the main screenshot of the interface that implements the \textit{guided} method. On the right is the detail for the specification of the acceptable range.}
\end{figure}

The \textit{guided-interaction} paradigm has been designed to take advantage of this algorithmic approach (see \cref{fig:PrefElic_figure}). 

\begin{enumerate}
    \item First, the user is provided an intervention that can overturn the decision given by the classifier;
    \item Second, the user has to rate the proposed intervention with a 5-point Likert scale ("Terrible", "Bad", "Neutral", "Good" and "Great");
    \item Then, for each feature modified by the intervention, the user can specify how achievable such change is on a 5-point Likert scale ranging from "Very difficult" to "Very easy". Users can also specify a preferred set of options for categorical features or an acceptable range for numerical features. See \ref{fig:guided-interface-control-tab}, left, for an example of the preference panel;
    \item Lastly, the user can decide to accept the proposed intervention or ask to generate a new one.
\end{enumerate}
 
\subsection{The control interface}
\begin{figure}[h]
  \centering
  \includegraphics[width=0.40\textwidth]{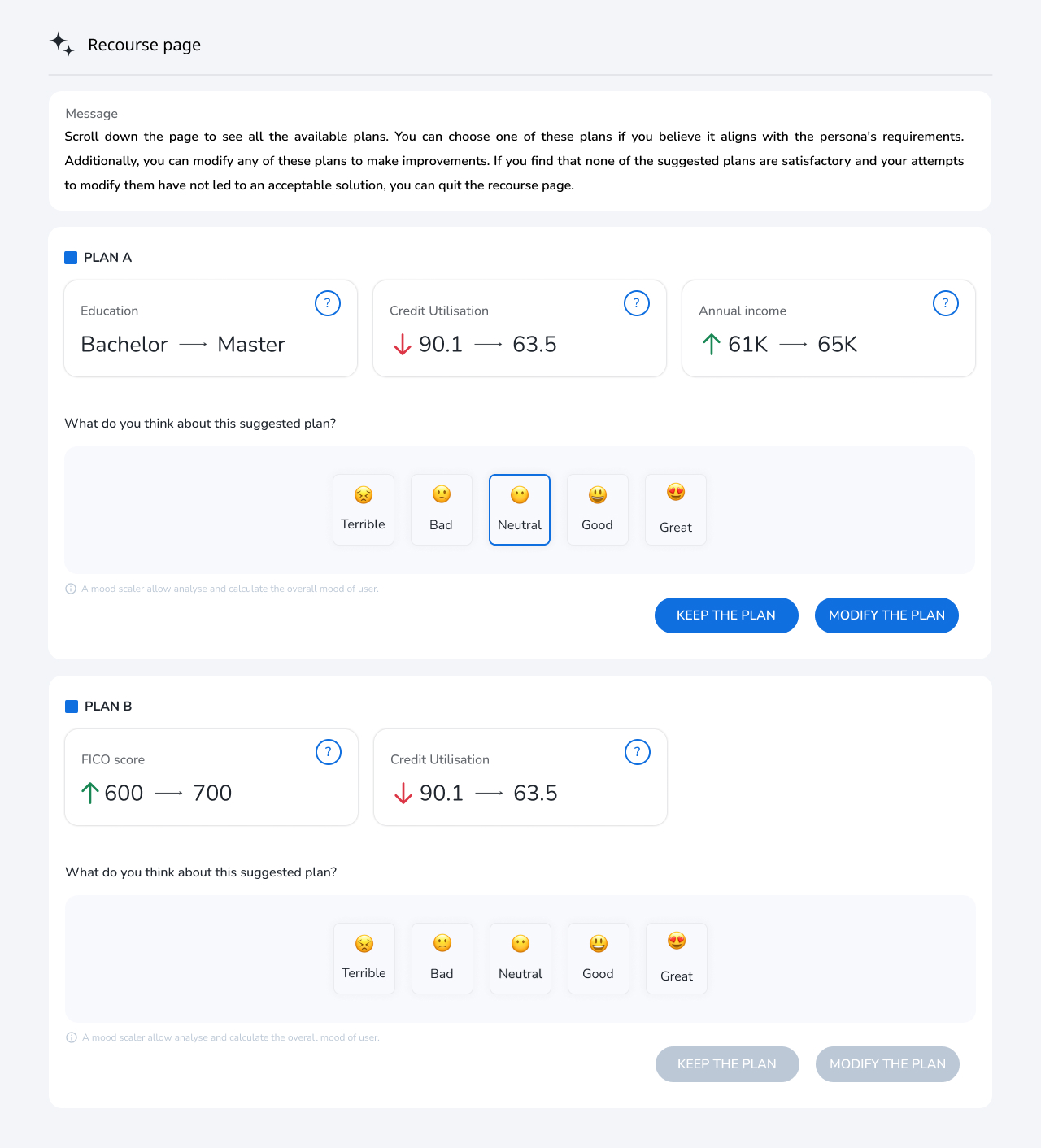}
  \includegraphics[width=0.40\textwidth]{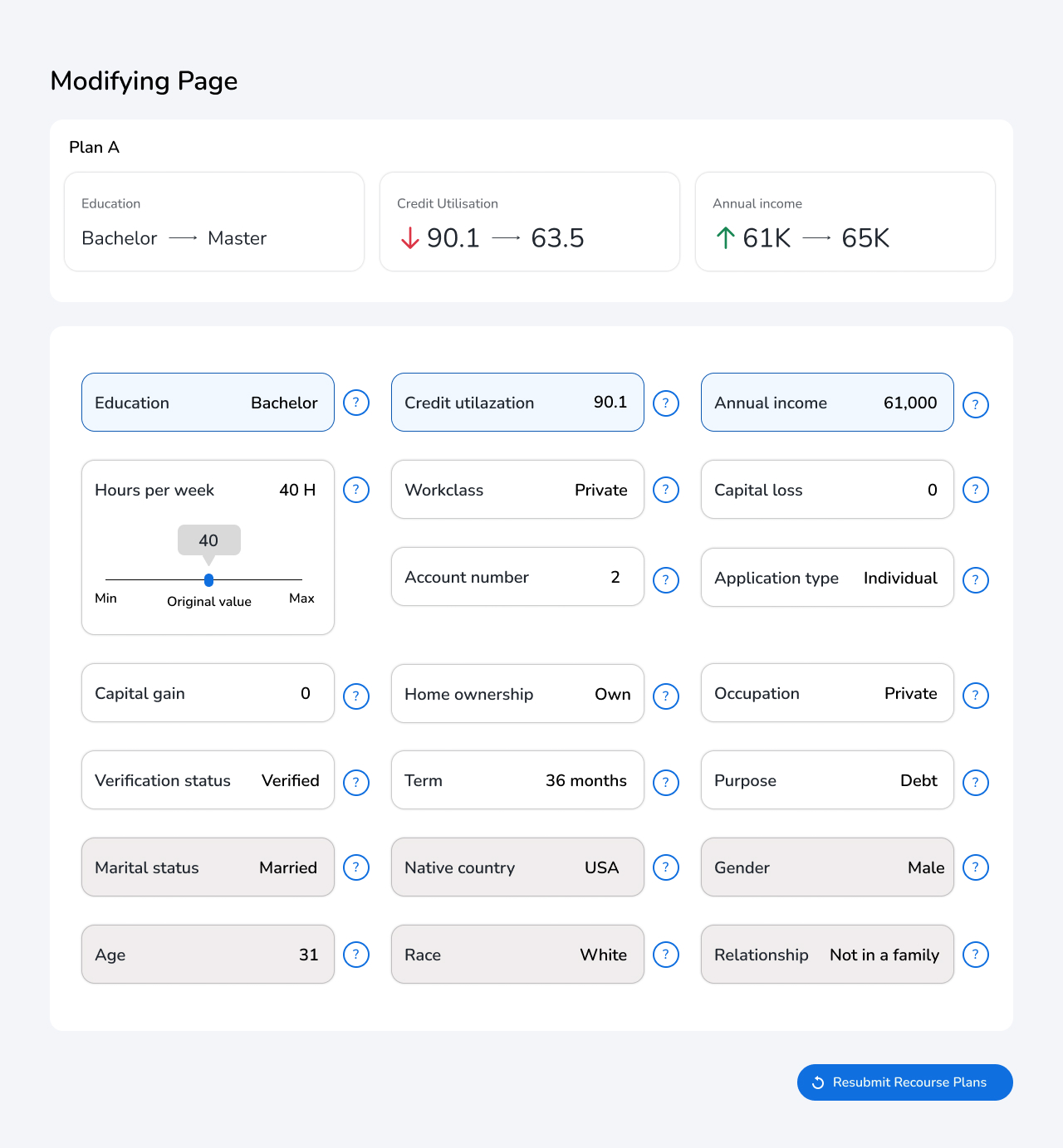}
  \caption{Control interface used in the study; inspired by GAM Coach \cite{wang_gam_2023}: the system proposes one or more plans, and the users can express their attitude and constraints on any available feature. On the left, two plans each one offering the possibility of accepting it or modifying it; on the right, one of the plans while being modified, all the actionable features can be accessed.}
   \label{Control_interface}
  \Description{Two screenshots of the interface used as the control condition in the study.}
\end{figure}

The graphical interface designed as a control condition of the study has been inspired by GAM Coach \cite{wang_gam_2023} but with a \textit{look-and-feel} closer to our interface discussed above. This has been done to reduce possible biases due to non-relevant aspects. The control interface implements a \textit{purely exploratory} perspective: one or more plans are generated, possibly considering previous constraints on features set by the user, but without learning from the user preferences. On the other hand, the interface allows the user to progressively disclose all the features and specify constraints on features not present in the suggested plan(s). Figure \ref{Control_interface} shows the main screen of the control interface.

\section{The Pilot Study}
For the pilot study, a loan application task has been chosen as a typical scenario for recourse assessment. In order to increase the number of features on which to base the decisions, two datasets: the "Adult Census Income" \cite{misc_adult_2} and the "Lending Club"\footnote{https://www.kaggle.com/datasets/wordsforthewise/lending-club} have been combined. Each dataset contains both continuous and categorical features for a total of 104 attributes. We selected a subset of 31 attributes that we believe the user would be reasonably capable of acting upon (such as, Job Category and Education Level). Then, we defined an action set $\calA$ comprising all the features, those 31 features were marked as \textit{actionable}, and all the others as \textit{non-actionable} but still used in the machine learning process. 

As the black-box classifier $h$, we trained an ensemble composed of two multi-layered perceptron architectures by minimizing the empirical risk over the training data via stochastic gradient descent \cite{ruder2016overview}. We trained the classifier to be as realistic as possible by using both the actionable and unactionable user attributes. In our loan scenario, we assume a loan is granted to a user only if both ensemble components give a positive classification.

\textbf{Study design.} The study employed a \textit{within-subjects} design. Each participant was asked to solve the same task (overturn an unfavorable outcome) with the \textit{guided}-style interface and with the \textit{purely exploratory}-style interface as a control condition. The order of the interfaces was randomized. 

\textbf{Procedure.} To make sure that the first interaction with the systems resulted in a rejection of the loan request (which was needed to start the recourse process), we used an approach based on \textit{personas}: fictional characters for which the specific characteristics were prepared in advanced \cite{matthews_how_2012}. While \textit{personas} are commonly used by designers to describe and empathize with users, we adopted this approach in user testing by asking our users to play the role of a specific \textit{persona} rather than responding on the interface for themselves. We also believe that this approach had the advantage of minimizing the potential influence of previous experience, personal knowledge, and attitudes toward lending. We defined two \textit{personas}, one female and one male. %
Each participant was asked to role-play one \textit{persona} while using one interface and the other \textit{persona} while using the other one. The association between interfaces and \textit{personas} was also randomized.

\textbf{Measures}. We adopted a mixed methods approach employing quantitative and qualitative measures. We employed the short version of the User-Experience Questionnaire (UEQ) \cite{
schankin_psychometric_2022, schrepp_design_2017}. This questionnaire consists of 26 semantic differential items on six scales: \textit{Attractiveness}, which measures the overall impression; \textit{Perspicuity}, which measures the ease of use and ease of learning, as well as, perceived familiarity; \textit{Efficiency}, how efficient and fast is perceived the product; \textit{Dependability}, how much the user feel in control of and safe during the interaction; \textit{Stimulation}, how exciting and motivating is to use the product; and finally \textit{Novelty}, that measures the degree to which the product is considered innovative and creative, and how much it captures users' attention. Since the questionnaire was administered by one experimenter, we selected one time for each scale, using the following anchors: \textit{Attractiveness}: unattractive vs attractive; \textit{Perspicuity}: complicated vs easy; \textit{Efficiency}: inefficient vs efficient; \textit{Stimulation}: boring vs exciting; \textit{Novelty}: conventional vs inventive; \textit{Dependability}: obstructive vs supportive. The experimenter administered the questionnaire at the end of the task with each interface. Besides scoring the semantic differential of each item, the experimenter also asked for a brief motivation. Those verbal reports were then transcribed and analyzed with Thematic Analysis \cite{liamputtong_thematic_2019}.  

\textbf{Participants.}  The call for participants was promoted among local university students, and the sampling was organized through a \textit{snowball effect}. Ultimately, 12 participants, three for each combination of \textit{persona}-interface, were recruited. The participants had an average age of 25.6 years (standard deviation 4.3; minimum 19 and max 33 years). Among them, there were ten females and two males. 

\section{Results and Discussion}

\begin{figure}[!h]
  \centering
  \includegraphics[width=0.6\textwidth]{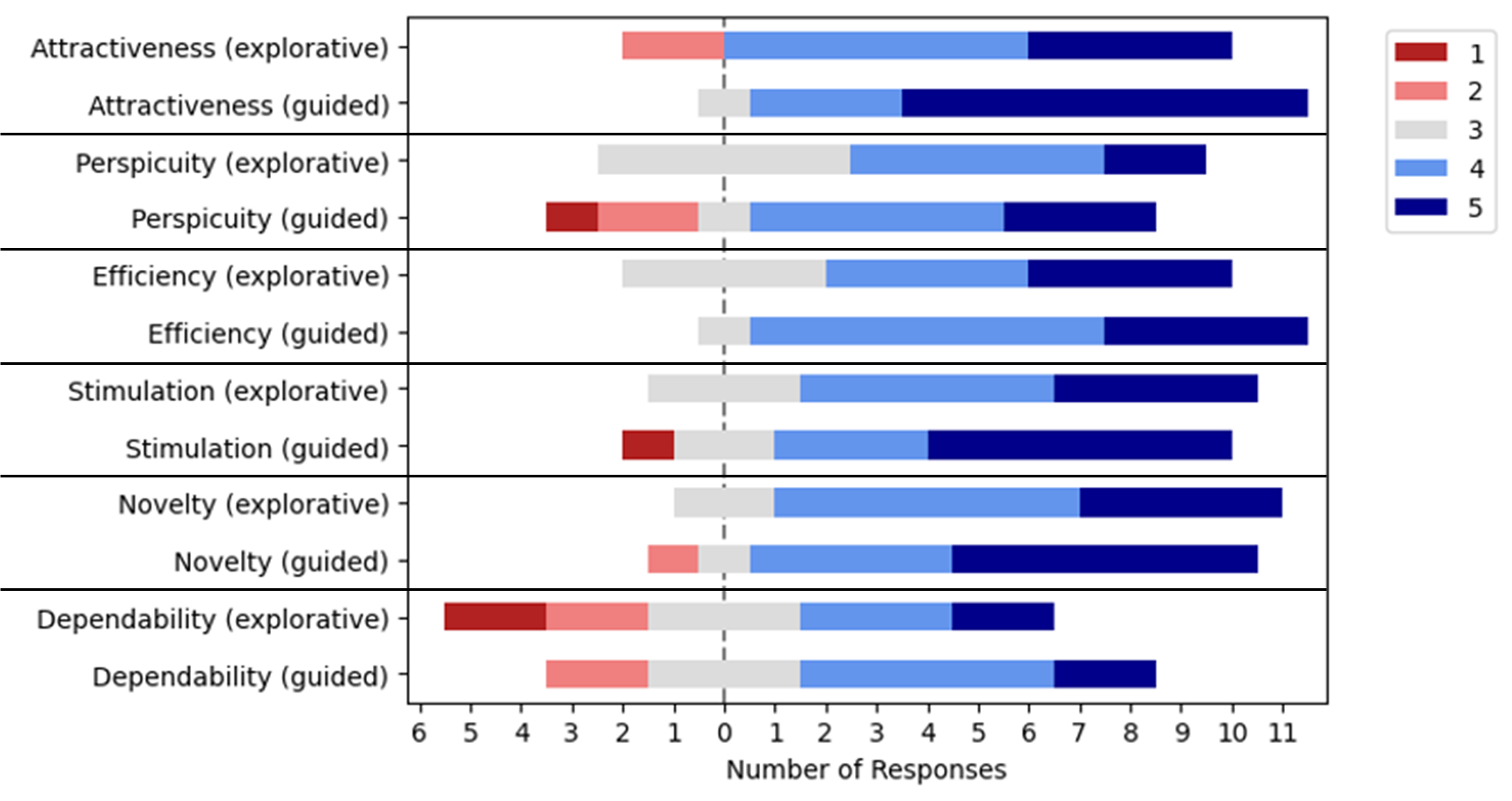}
  \caption{Scores of the UEQ questionnaires for the two interfaces: the value 1 corresponds to the negative anchor and the value 5 corresponds to the positive anchor for the semantic differential items.}
   \label{UEQ_questionnare_figure}
\end{figure}

Figure \ref{UEQ_questionnare_figure} presents the scores of the semantic differentials. The overall scores were good with an average above 4, except Perspicuity and Dependability, whose scores ranged between 3 and 4. None of the observed differences are statistically significant, as confirmed by Friedman's ranking tests (with Bonferroni's correction).

For the \textit{guided}-style there is a positive correlation between the duration of the task and the dimension of Attractiveness (Pearsons' ro=2.6), Perspicuity (ro=0.73) and Dependability (ro=0.15) while the same correlations are negative for the \textit{exploratory}-style interface (respectively, -0.23,	-0.32,	and -0.12); for the Efficiency score, the correlation is negative for the  \textit{exploratory}-style interface and not statistically different from 0 for the \textit{guided}-style interface. Stimulation does not seem to correlate with duration in either one of the interfaces. While these results are preliminary due to the low number of participants, it's noteworthy that our users appreciate the freedom provided by the \textit{explorative}-style interface, unless it leads to time wastage. Conversely, more time spent on the \textit{guided}-style interface is perceived as useful, possibly because it offers a more direct path to finding a solution.

From a Thematic Analysis of the verbal reports collected as motivations for the scores of the scales, it emerges that our participants articulated their experiences along two different dimensions. On one side, the freedom to propose changes and manipulate the input (even though our participants never engaged in such elaborate behaviors). On the other side, the need to avoid confusion and quickly reach a solution. 

For example, p1011 commenting on Perspicuity for the exploratory interface said \textit{“I could solve them without effort”} and in commenting Efficiency for the guided \textit{“I barely put effort for this”}. Yet, in commenting on Stimulation, she said \textit{“I was more excited to use this one [the exploratory] because I had more plans to check. I liked it more.”} Nevertheless, for the guided interface, she reflected that \textit{“Maybe for someone that wants loan approval in real life, it would be more interesting.”}. 
P1013 discussing Attractiveness said \textit{“I prefer the first platform [the guided-style], in the second one [the exploratory] I could express my preferences with more details but it took too much time and confused me.”}
P1016 on commenting the Perspicuity of exploratory “\textit{it was harder than the first one, but this one you had more options to change I prefer this one.”} She was also more explicit on commenting Stimulation for the exploratory style \textit{“[this one is] more exciting than [the one] before because I had more control over changing. The fact that you have more plans is more exciting than having one plan.”}
On a similar line, P018 on Stimulation of the exploratory \textit{"this one is more exciting [...] you could change the plan then you have more plans again.”}
P1014 on commenting on the Dependability of the exploratory summarized these terms: \textit{“I didn't expect to receive many plans [positive surprise], and negative surprised because of a long list of information in the modifying page [the list of all the disclosed features].”}.
In a similar venue, P1019 on the Efficiency of the exploratory \textit{“I think it was 5, but the other one was shorter, and people (may) find the other one more efficient.”}.

\section{Conclusion}
Our work addressed a core aspect of the design of \textit{Human-Centred AI}, namely, how to balance user control with system autonomy \cite{shneiderman2022human}, in the context of AI-based algorithm recourse. Our design was framed in the strategy of \textit{prospective interaction}. Still, it differentiates from a \textit{purely exploratory} approach by framing the exploration within the boundaries of an optimized preference elicitation process.

The aim of our work was not just to compare our interface to GAM Coach  \cite{wang_gam_2023} but rather to compare the two approaches of \textit{prospective} methods for AI-driven systems: a purely \textit{exploratory} approach and a purely \textit{guided} approach. It is worth noting that they differ concerning (i) how the algorithmic recourse is realized in the backend and (ii) how the interface fosters either free interactions up to what-if explorations or rather an efficient but constrained step-by-step process. However, in both cases, users are engaged in proactively "probing the algorithm boundaries with different inputs" \cite{shneiderman2022human}.  
Our results suggest that users may recognize that the \textit{guided} interaction paradigm improves efficiency. However, they also feel less freedom to experiment with "what-if" scenarios. Nevertheless, the time spent on the  \textit{purely exploratory} interface tends to be perceived as a lack of efficiency, which reduces attractiveness, perspicuity, and dependability. Conversely, for the \textit{guided}-style interface, more time on the interface seems to increase its attractiveness, perspicuity, and dependability while not impacting the perceived efficiency. The evidence of this study might suggest that this type of interfaces should combine these two approaches by trying to support \textit{exploratory} behavior while gently pushing toward a \textit{guided} effective solution. 

We acknowledge that the main limitations of our study include the small sample size and the likelihood that our participants may not have fully grasped the task's difficulty due to their lack of previous experience in money lending. Nevertheless, albeit still preliminary, we believe our work may help shed new light on the ongoing debate on \textit{Human-Centred AI} and lay the basis for further studies on this topic.

\begin{acks}
The work was partially supported by the following projects:  AI@Trento, FBK-Unitn (GDT),  MUR PNRR project FAIR - Future AI Research (PE00000013) funded by the NextGenerationEU  (AP, BL and MZ)
 Horizon Europe Programme, grant number 101120237 - ELIAS (AP and BL) and grant number 101120763 - TANGO (AP, BL and KT). 
Funded by the European Union. Views and opinions expressed are however those of the author(s) only and
do not necessarily reflect those of the European Union or the European Health and Digital Executive Agency
(HaDEA). Neither the European Union nor the granting authority can be held responsible for them. 

\end{acks}

\bibliographystyle{ACM-Reference-Format}
\bibliography{main}

\end{document}